\documentclass{jfm}

\usepackage{graphicx}
\usepackage{epsfig}
\usepackage{amsmath}
\usepackage{amssymb}
\usepackage{amsfonts}

\newcommand{\be}{\begin{equation}}
\newcommand{\ee}{\end{equation}}
\newcommand{\bea}{\begin{eqnarray}}
\newcommand{\eea}{\end{eqnarray}}
\newcommand{\bc}{\begin{center}}
\newcommand{\ec}{\end{center}}
\newcommand{\btab}{\begin{tabular}}
\newcommand{\etab}{\end{tabular}}

\let\oldepsilon\epsilon
\let\epsilon\varepsilon
\let\varepsilon\oldepsilon
\let\oldphi\phi
\let\phi\varphi
\let\varphi\oldphi
\renewcommand{\deg}{{}^{\circ}}

\title{Relaxation of a dewetting contact line\\ Part 2: Experiments}

\author[G. Delon, M. Fermigier, J.H. Snoeijer and B. Andreotti] {G\ls I\ls L\ls E\ls S\ns  D\ls E\ls L\ls O\ls N$^1$,\ns M\ls A\ls R\ls C\ns  F\ls E\ls R\ls M\ls I\ls G\ls I\ls E\ls R$^1$,\ns \\ J\ls A\ls C\ls C\ls O\ns H.\ns S\ls N\ls O\ls E\ls I\ls J\ls E\ls R$^{1,2}$\ns \and B\ls R\ls U\ls N\ls O\ns A\ls N\ls D\ls R\ls E\ls O\ls T\ls T\ls I$^1$\ns}

\affiliation{$^1$ Physique et M\'ecanique des Milieux 
H\'et\'erog\`enes, ESPCI, 10 rue Vauquelin, 75231 Paris  
Cedex 05, France \\
$^2$ School of Mathematics, 
University of Bristol, University Walk,
Bristol BS8 1TW, United Kingdom}
\date{\today}
\pubyear{2000}
\volume{???}
\pagerange{??--??}
\date{\today}
\begin{document}
\maketitle

\begin{abstract}
{The dynamics of receding contact lines is investigated experimentally through controlled perturbations of a meniscus in a dip coating experiment. We first characterize stationary menisci and their breakdown at the coating transition. It is then shown that the dynamics of both liquid deposition and long-wavelength perturbations adiabatically follow these stationary states. This provides  a first experimental access to the entire bifurcation diagram of dynamical wetting, confirming the hydrodynamic theory developed in Part 1. In contrast to quasi-static theories based on a dynamic contact angle, we demonstrate that the transition strongly depends on the large scale flow geometry. We then establish the dispersion relation for large wavenumbers, for which we find that $\sigma$ is linear in $q$. The speed dependence of $\sigma$ is well described by hydrodynamic theory, in particular the absence of diverging time-scales at the critical point. Finally, we highlight some open problems related to contact angle hysteresis that lead beyond the current description.}
\end{abstract}

\section{Introduction}
Moving contact lines have been studied for more than thirty years but constitute still an open problem in fluid mechanics. The difficulty comes from the existence of six decades of length scale separating the macroscopic scale from the molecular scale that become active as soon as a contact line moves, due to viscous diffusion. This effect may be seen in the classical hydrodynamics description, where the no-slip boundary condition leads to a divergence of viscous stresses  at the contact line (\cite{HS71, DDD74}). Of course, this singularity can be avoided by considering molecular physics that goes beyond hydrodynamics, such as the description of diffuse interfaces (\cite{PP00}), Van der Waals interactions (\cite {TDS88}), or a slip at the solid substrate (\cite{TR89}). The latter mechanism has recently been accessed experimentally (\cite{SHL05,CCSC05}), showing that slip really occurs and is not an ad hoc quantity to save the hydrodynamic description. Over a large range of shear rates, the velocity $v_s$ of the last layer of molecules was found proportional to the velocity gradient $\dot \gamma$,
\begin{equation}
v_s=l_s \dot \gamma~,
\end{equation}
where $l_s$ is the slip length. According to these experiments and molecular dynamics simulations (\cite{TT97,BB99}), large slip lengths are associated to a hydrophobic behaviour. For moderately large contact angles, the slip length is of the order of a few molecule sizes. Even though, the difficulty of the moving contact line problem arises from the very large interface curvatures near the contact line, required to balance the viscous stresses (\cite{V76,C86}). This strongly curved region has to be matched to the macroscopic flow, which is particularly challenging in the dewetting case (\cite{E04, E05}).

On the experimental side, this problem is essentially studied by examining the macroscopic interface shape as a function of the properly rescaled contact line speed $U$, (e.g. see \cite{H75, DRG91, LG05}), called the capillary number:
\begin{equation}
{\rm Ca}=\frac{\eta U}{\gamma}~,
\end{equation}
where $\eta$ and $\gamma$ are viscosity and surface tensions respectively. 
However, macroscopically observable parameters, such as the dynamic contact angle, are not very sensitive to distinguish the microscopic contact line models. Golestanian \& Raphael proposed that, by studying {\em perturbations} of contact lines, one could discriminate between different dissipation models at the contact line. Their analysis is based on  the elastic-like description for \emph{static} contact lines (\cite{JdG84, DG86a}): a small perturbation of the contact line position with wavector $q$ involves the deformation of the free surface over a distance $1/q$ resulting in an elastic capillary energy proportional to $|q|$. The contact line returns to its equilibrium straight configuration with a characteristic time $\sigma^{-1}$ such that, in the limit of small contact angles $\theta$,
\begin{equation}
\sigma \propto \frac{\gamma}{\eta} \theta^3 |q| .
\label{eq:scalingpgg}
\end{equation}
The $\theta^3$ dependence reflects the visco-capillary balance within the wedge of liquid bounded by the solid substrate and the free surface. \cite{OV91b} were the first to experimentally study this dispersion relation for a static contact line and they confirmed in particular the $|q|$ dependence in the limit of large $q$. \cite{MC93a} examined the relaxation of a very slowly moving contact line, distorted by an isolated chemical defect. They showed that the relaxing line profiles can be described by functions of the form $\ln(y^2 + c^2t^2)$, where $y$ is the coordinate along the contact line and $c$ is the characteristic speed $\propto \gamma \theta^3/\eta$. This logarithmic shape is also a direct consequence of the peculiar contact line elasticity (\cite{DG86a}). 

In the case of \emph{receding} contact lines, the quasi-static theory by Golestanian \& Raphael predicts that the relaxation time $\sigma^{-1}$ should increase with contact line speed and diverge at the dynamic entrainment transition, i.e. when a steady meniscus can no longer be sustained. An intriguing consequence of this is that perturbations due to small-scale inhomogeneities of the substrate are no longer damped at the critical point, leading to a roughening of the contact line (\cite{GR03}). This scenario contrasts the dispersion relation obtained from the full-scale hydrodynamic calculation presented in our preceding paper (Part 1, \cite{SADF07}), predicting a finite relaxation time for perturbations smaller than the capillary length. This hydrodynamic calculation explicitly accounts for viscous dissipation at all lengths and is thus expected to be more accurate than a quasi-static approach, in which dissipative effects enter through an effective boundary condition.

In this paper we experimentally study the global stability and relaxation times of a contact line in the context of a simple dip-coating experiment (figure~\ref{fig:setup}). When a vertical plate is withdrawn from a liquid bath at velocities below the coating transition, the contact line equilibrates and we study the relaxation of well-controlled perturbations. It is found that the relaxation times indeed increase as the entrainment transition is approached. However, as we have shown previously (\cite{SDFA06}), the transition is not critical because relaxation times remain finite at threshold. The full dispersion relation is established and compared quantitatively to hydrodynamic results. Above the transition it is found that transients evolve adiabatically through a succession of quasi-steady states. We can thus, for the first time, experimentally access the full bifurcation structure of the wetting transition, using these transient states. Our experiments confirm the nontrivial bifurcation scenario proposed in Part 1. 

The paper is organized as follows. In Sec.~\ref{sec:setup} we describe briefly the experimental set-up and the physico-chemical properties of the system used. The framework of the hydrodynamic theory developped in Part 1 is briefly recalled in Sec.~\ref{sec:hydro}. In Sec.~\ref{sec:steady}, we then examine the global shape of the meniscus, essentially characterized by its height above the liquid bath. We determine the critical velocity for meniscus stability and investigate the bifurcation diagram from transients evolution to liquid deposition. Section~\ref{sec:dispersion} is devoted to the analysis of the contact line relaxation. We first examine periodic pertubations created by rows of defects moving through the contact line. These perturbations are shown to decay with a rate $\sigma$ proportional to the wavevector $q$, as for a static contact line. We also examine the $q=0$ mode, i.e. the relaxation of the average meniscus height to its stationary position. In Sec.~\ref{sec:results} we show that the variation of $\sigma$ with respect to the capillary number and its behavior near the entrainment transition are well described by the hydrodynamic theory. We complete this discussion of contact line relaxation, in Sec.~\ref{sec:lorentz}, by presenting experiments on localized perturbations. In the conclusion we finally address several open problems in contact line dynamics, particularly, the possible influence of hysteresis which has not yet been studied properly.

\section{Experimental set-up}\label{sec:setup}
\begin{figure}
\includegraphics{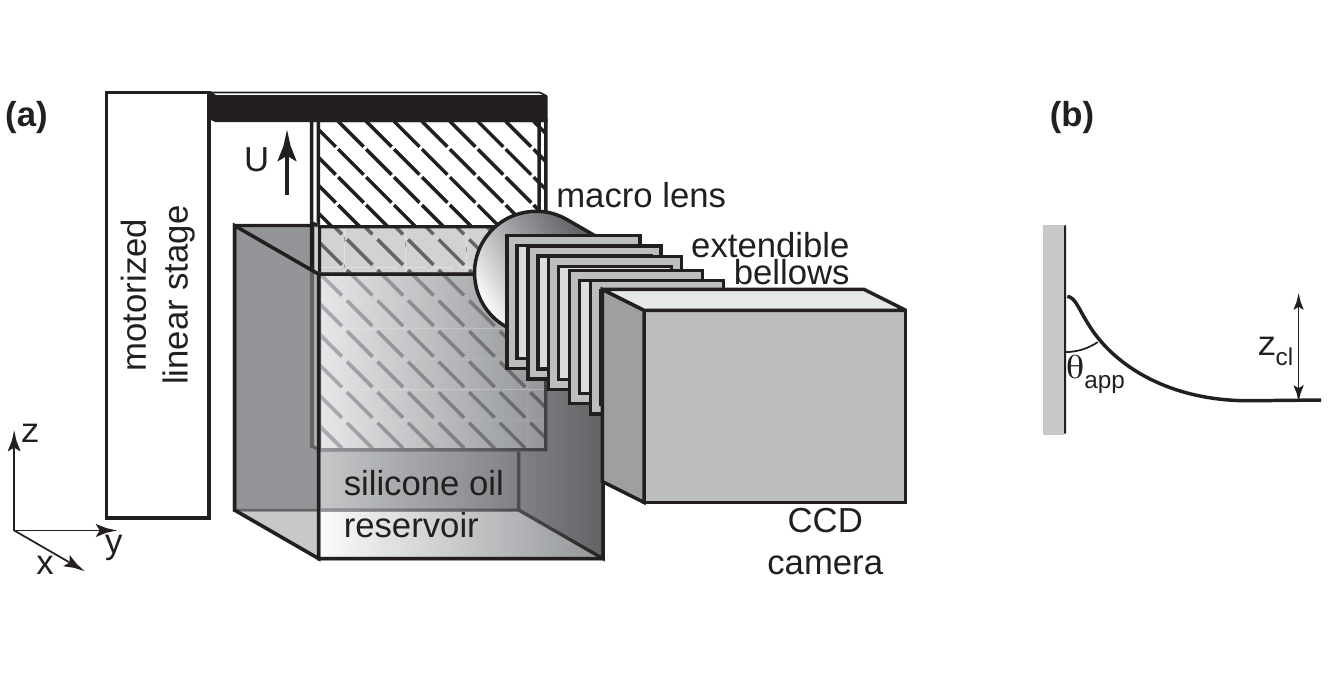} 
\caption{Experimental set-up. {\em (a)} A vertical plate is withdrawn at velocity $U_p$ from a bath of liquid that does not wet spontaneously on it. {\em (b)} Definition of meniscus rise $z_{\rm cl}$ and the apparent contact angle $\theta_{\rm app}$.} 
\label{fig:setup}
\end{figure} 

The experiment simply consists of withdrawing a non-wetting plate from a vessel filled with viscous liquid (figure~\ref{fig:setup}). The plate is a $5~$cm wide strip, cut from a silicon wafer (Siltronix). A thin layer of fluorinated material is deposited on the wafer by dip coating in a solution of FC 725 (3M) in ethyl acetate. The liquids used are polydimethylsiloxanes (PDMS, Rhodorsil 47V series) with dynamic viscosities $\eta$ ranging from $1$ to $5~$Pa.s (the corresponding average molecular weights range from 21000 to 40000), surface tension $\gamma=21$~mN/m and density $\rho=980~\mathrm{kg/m^3}$. The corresponding capillary length is $l_{\gamma}=\sqrt{\gamma/\rho g} =1.46$~mm. This particular physico-chemical system was chosen because high molecular weight PDMS is non volatile and its low surface tension inhibits rapid contamination of the free surface. In addition, this allows a direct comparison with other experiments performed with the same system in a different geometry.

PDMS is a molten polymer and it exhibits an entanglement transition  at a molecular weight around 20000 (\cite{RLHBHSNP84}). The flow behavior is Newtonian up to a critical shear rate $\dot{\gamma}_c$ which decreases with the molecular weight. For the fluids used in this study, $\dot{\gamma}_c \approx 10^4 \, \mathrm{s}^{-1}$ (\cite{LPK70}). This critical value, above which shear thinning is observed, should be compared to the experimental shear rates at the macroscopic and microscopic scales. At the macroscopic scale $\dot{\gamma} \approx U_p/l_{\gamma}$, which never exceeds 0.1 $\mathrm{s}^{-1}$. Thus we expect a purely newtonian behavior of the liquid at the scale of the capillary length. At the microscopic scale $\dot{\gamma} \approx U_p/a$, where $a$ is a molecular size of the order of 10 nm. The shear rate can thus reach $10^4\, \mathrm{s}^{-1}$ very close to the contact line and a moderate decrease of the viscosity might take place (\cite{LPK70}).  

We were not able to measure directly the slip length of our system, but it can be estimated as follows. Starting from the length of the Si-Si binding (around $0.3$~nm) and from the number of monomers (around $25^2$ for the high viscosity oil of $\eta=4.95\,$ Pa.s), we obtain the size $a\simeq 7.5$~nm of a molecule (\cite{LG05}). It is known from molecular dynamics simulations that contact angles lower than $90\deg$, for which the interaction between the liquid and the substrate is attractive, give rise to a slip length of the order of 2 molecular lengths (\cite{TT97}). Throughout the paper we therefore use the value $l_s\simeq 2a\simeq 15$~nm$\simeq 10^{-5}~l_{\gamma}$ when comparing to theoretical results.

PDMS partially wets the fluorinated coating with a static contact angle that can vary from one plate to another by $5\deg$. The data presented here have been obtained for a receding contact angle of $\theta_r=51.5\deg$ and an advancing contact angle of $\theta_a=57.1\deg$. Like all the plates prepared for this study, the contact angle hysteresis is thus very low ($\theta_a-\theta_r<7\deg$), as previously obtained (\cite{RDAL05}).

To induce controlled perturbations of the contact line we create wetting defects on the plate using two techniques:
\begin{itemize}
\item controlled deposition of ink droplets on the fluorinated coating. When dried, ink  has a much higher surface energy than the fluorinated coating, and it is completely wetted by the silicone oils.
\item spin-coating a layer of photo-sensitive resin (SU-8 Microchem) on the surface of a silicon wafer. After UV exposure through a mask the resin is developed, leaving cylindrical posts ( 200 $\mu$m wide, 100 $\mu$m high) on the wafer. The whole surface is then coated with FC-725, as described above. With this technique, the surface wettability is uniform and the defects are only physical.
\end{itemize}
Both fabrication techniques produce surface defects that are able to significantly distort the contact line as they move through the meniscus. 

The size of the vessel containing the liquid is chosen sufficiently large (10 $\times$ 10 $\mathrm{cm}^2$) to avoid any capillary interaction between the meniscus on the plate and the menisci formed on the rim of the vessel. Also, the cross section of the silicon wafer is $10^{-3}$ times the cross section of the vessel, so that the liquid displacement by the wafer hardly affects the vertical position of the free surface. When the plate moves at its typical high velocity, 100 $\mu$m/s, the reference level in the bath is displaced only at 0.1 $\mu$m/s.

The motion of the plate is controlled within $1~\mu$m by a motorized linear stage (Newport Corp., linear stage M-UTM50, controller ESP300). The image of the meniscus is recorded with a CCD camera (Basler A602f, 656x492 pixels, pixel size: $9.9~\mu$m x $9.9~\mu$m, $100$~frames/s) fitted with a macrophotography bellows and a Nikon $2.8/60~$mm lens. We can thus obtain a magnification ratio of $5$, in which case $1$ pixel in the image corresponds to $2~\mu$m on the object plane. 

The location of the contact line is precisely determined  by a cross-correlation procedure. The gray level profile corresponding to the unperturbed contact line is recorded for each experiment. This reference profile is then correlated with each vertical line of the image. The contact line position corresponds to the location of the correlation maximum. The location of this maximum is subsequently refined with subpixel resolution by interpolation around the correlation peak. This procedure is implemented as a plugin for the ImageJ software (http://rsb.info.nih.gov/ij/).

\section{Hydrodynamic framework}
\label{sec:hydro}
Let us briefly describe the hydrodynamic theory to which the experimental results will be compared. We basically follow the analysis of the accompanying paper, Part 1, \cite{SADF07}, which is based upon the lubrication approximation for noninertial free surface flows~(\cite{ODB97,H01,E04}). However, to enable a {\em quantitative} comparison involving large contact angles, typically around $45^\circ$, we include corrections to the standard lubrication theory as proposed by \cite{S06}. The governing equations for the interface profile $h(z,y,t)$ then become 
\begin{eqnarray}
\label{continuity}
\partial_t h+ \nabla \cdot \left( h \,{\bf U}\right) &=& 0~, \\
\nabla \kappa - {\bf e}_z + \frac{3({\rm Ca}\,{\bf e}_z  - {\bf U})}{h(h+3l_s)} F(\theta) &=& {\bf 0}~,
\label{momentum}
\end{eqnarray}
representing mass conservation and force balance respectively. Here, ${\bf U}$ is the depth-averaged velocity, while $\kappa$ is twice the mean curvature of the interface. The equations have been made dimensionless using the capillary length $l_\gamma$ and the capillary time $\eta l_\gamma /\gamma$. The equations differ from the standard lubrication approach through a correction factor 
\begin{equation}
F(\theta) = \frac{2}{3}\frac{\tan \theta \sin^2 \theta}{\theta-\cos \theta \sin \theta}~,
\end{equation}
where $\tan \theta$ is the local slope of the interface (\cite{S06}). Indeed, $F(\theta)\simeq 1$ for $\theta \ll 1$. We refer to Part 1 for details on boundary conditions and the numerics of the linear stability analysis.

The theory requires two input parameters characterizing the contact line: the slip length $l_s$, preventing a stress divergence, and a microscopic contact angle $\theta_{\rm cl}$. As argued in Sec.~\ref{sec:setup}, we can use a value $l_s =10^{-5}~l_{\gamma}$ estimated from the molecular size. Macroscopic results depend only logarithmically on the precise value of $l_s$ (\cite{V76,C86}). The microscopic contact angle is unknown a priori, but it is generally assumed to be equal to the equilibrium angle. For hysteretic systems, the static angle can take any value between $\theta_r$ and $\theta_a$. Since the results are quite sensitive to this parameter, we have produced numerical curves using three different values of $\theta_{\rm cl}$: receding angle $\theta_r=51.5^\circ$, advancing angle $\theta_a=57.1^\circ$ and average static angle $(\theta_{a}+\theta_r)/2=54.3^\circ$.

\section{Steady menisci}
\label{sec:steady}

\subsection{Contact line position as a function of capillary number}
\begin{figure} 
\includegraphics{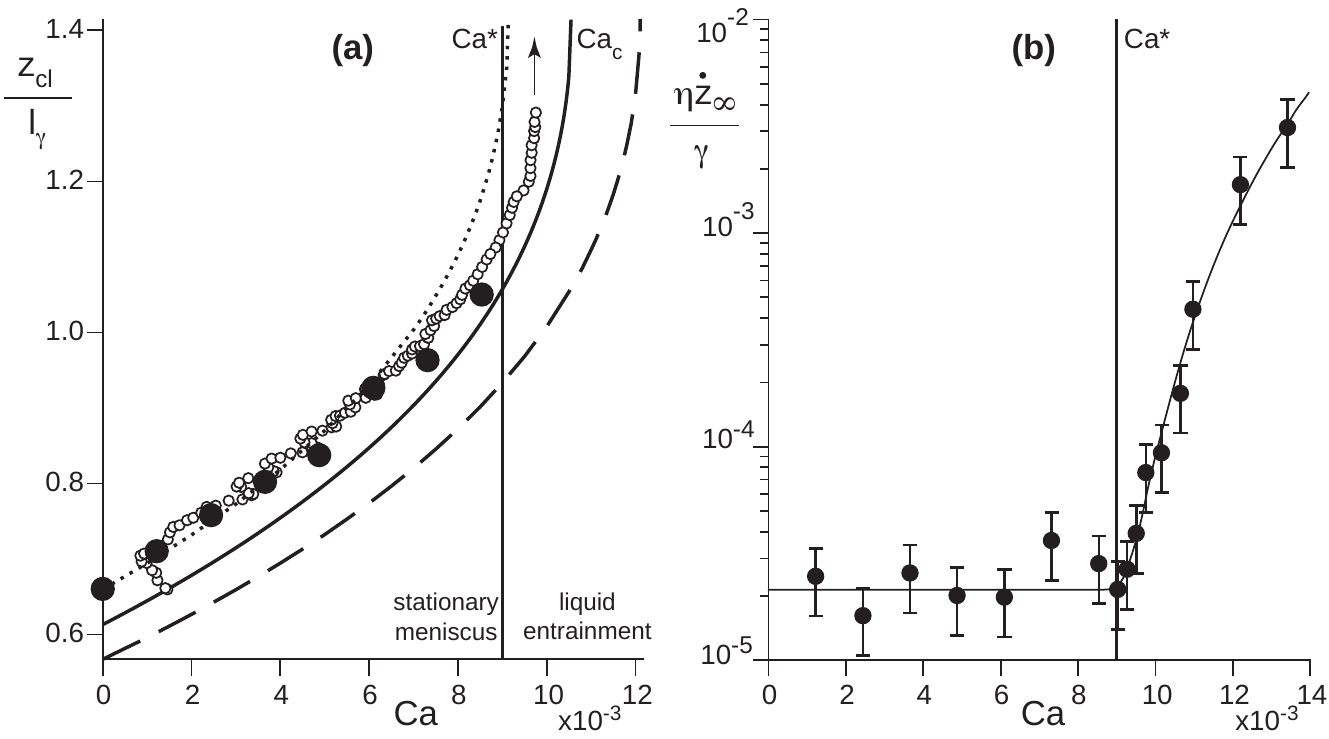} 
\caption{{\em (a)} Meniscus rise $z_{\rm cl}$ normalized by capillary length $l_{\gamma}$ as a function of the plate capillary number ${\rm Ca}$. Symbols $(\bullet)$: steady solutions, determined experimentally as a function of ${\rm Ca}$. Symbols $(\circ)$: rescaled meniscus rise $z_{\rm cl}(t)$ function of the contact line relative capillary number $\widetilde{\rm Ca}(t)=\eta (U_p-\dot{z}_{\rm cl}(t))/\gamma$, for ${\rm Ca}=9.8~10^{-3}$ (see text). Lines: predictions from hydrodynamics theory, with microscopic contact angle $\theta_{\rm cl}=\theta_r$ (dotted), $\theta_{\rm cl}=(\theta_a+\theta_r)/2$ (solid), $\theta_{\rm cl}=\theta_a$ (dashed). {\em (b)} Rescaled contact line velocity at long time $\dot{z}_{\infty}$, as a function of the capillary number  ${\rm Ca}$. Each point corresponds to an average over several experiments. The error bars indicate the typical variation of the measured quantity from one experiment to the other. The solid line is a phenomenological fit of the form: $\eta \dot{z}_{\infty} / \gamma=c_1+c_2 ({\rm Ca}-{\rm Ca^*})^3$. $c_1$ is a residual ascending velocity present even below the threshold.}
\label{fig:zcl} 
\end{figure}

When the vertical plate is at rest, the liquid rises above the bath up to a height $z_{\rm cl}$, figure~\ref{fig:setup}, determined by the capillary length and the contact angle, according to the classical relation (\cite{LL59}),
\begin{equation}
z_{\rm cl}=l_{\gamma}\sqrt{2(1-\sin\theta)}~,
\label{eq:zclvstheta}
\end{equation}
where $\theta$ is the equilibrium contact angle (receding or advancing). This relation implies that a perfectly wetting liquid can achieve a maximum rise of $\sqrt{2}$ times the capillary length $l_{\gamma}$. 

When the plate is set withdrawn with a velocity $U_p$, so that the contact line recedes with respect to the plate, the meniscus height increases to a new equilibrium value. The closed circles on figure~\ref{fig:zcl}a represent experimentally observed $z_{\rm cl}$ for various ${\rm Ca}=\eta U_p/\gamma$, showing an increase of the meniscus rise with ${\rm Ca}$. However, beyond a critical velocity, corresponding to a capillary number ${\rm Ca}^*$, the meniscus no longer equilibrates but rises indefinitely. This is the signature of the entrainment transition: in our experiments, steady menisci cannot exist beyond ${\rm Ca}^* \approx  9.1 \times 10^{-3}$. 

The dependence of $z_{\rm cl}$ on ${\rm Ca}$ can be compared to the predictions of hydrodynamic theory. As mentioned in Sec.~\ref{sec:hydro}, the numerical curves are quite sensitive to the boundary condition of the microscopic contact angle, $\theta_{\rm cl}$. In figures~\ref{fig:zcl}a we therefore present numerical curves obtained using $\theta_{\rm cl}=\theta_r$, $\theta_{\rm cl}=\theta_a$ and $\theta_{\rm cl}=(\theta_{a}+\theta_r)/2$. The experimental points for $z_{\rm cl}({\rm Ca})$ lie between the curves obtained with $\theta_r$ and the average static angle. It should be noted that, while we can measure the \textit{relative} contact line motion with a precision of a few microns, it is much more difficult to get the reference level of the liquid bath, inducing incertainty in the static angles. There is, however, an important discrepancy on the precise location of the transition: for all model parameters, the hydrodynamic theory predicts that the transition occurs when the meniscus reaches $z_{\rm cl}=\sqrt{2}l_\gamma$, the height attained by a perfectly wetting liquid (Part 1, \cite{E04}). We denote this theoretical maximum velocity as the critical point, with a critical capillary number ${\rm Ca}_c$. In the experiments, entrainment already occurs at $z_{\rm cl}\approx 1.1 l_\gamma$, from which we infer that ${\rm Ca}^* < {\rm Ca}_c$. Below we discuss how the experimental ${\rm Ca}^*$ is related to transient film solutions.

\begin{figure} 
\includegraphics{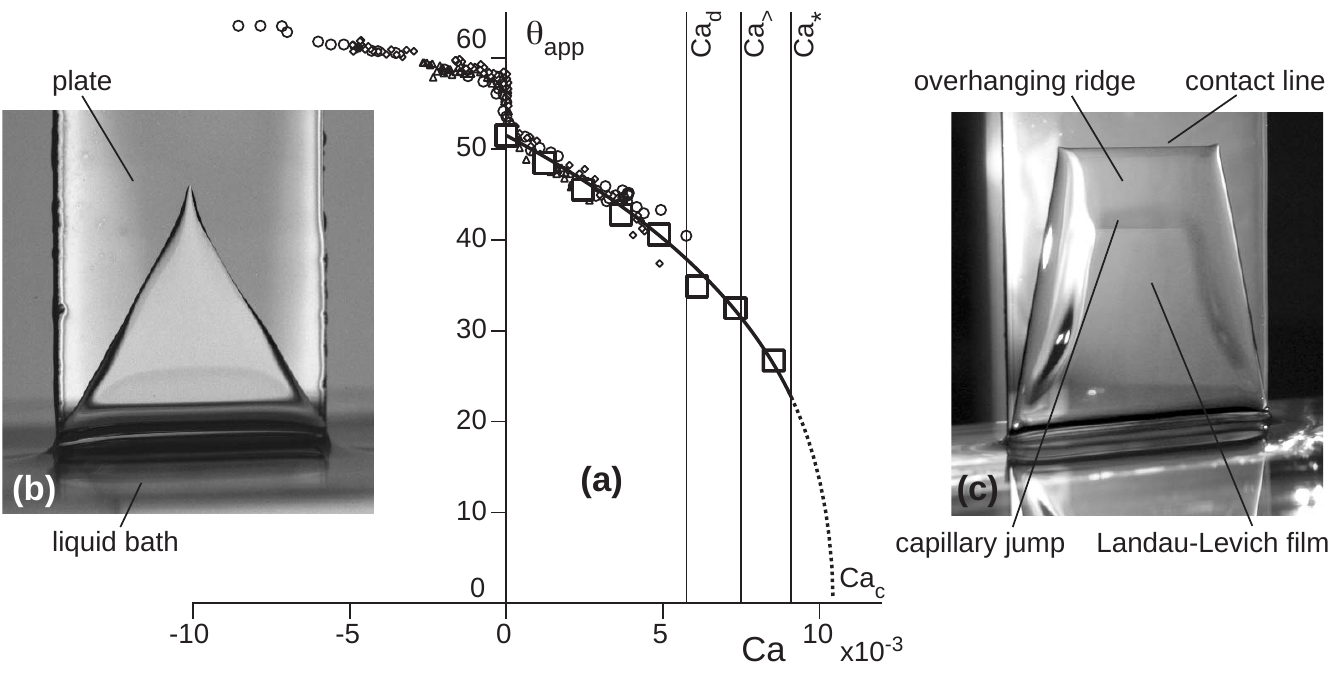} 
\caption{{\em (a)} Symbols $(\square)$: apparent contact angle $\theta_{\rm app}$, defined from Eq~(\ref{eq:thetaapp}) as a function of capillary number for PDMS on fluorinated glass or silicon. Symbols ($\Large{\circ}$, $\Large{\diamond}$, $\triangle$): data from \cite{RDAL05} obtained for drops sliding on an inclined plane. Vertical lines indicate threshold capillary numbers ${\rm Ca}_d$ for the drop experiment, ${\rm Ca}_>$ "corner" dewetting flow on vertical plate (see text), ${\rm Ca}^*$ for the entrainment transition in the plate geometry. The solid line is the result of the hydrodynamic theory for $(\theta_{a}+\theta_r)/2$, shifted down by $2.8^\circ$. It is mostly used as a guide eye but shows that the prediction is within the experimental error on the absolute position of the contact line. {\em (b)} Triangular liquid film observed when the dewetting lines originating at the wafer edges meet. {\em (c)} Overall shape of the liquid film well above the entrainment transition. Most of the analysis pertains to the horizontal contact line at the top of the film.} 
\label{fig: ThetaCa} 
\end{figure} 

These results can be represented in terms of the {\em apparent} contact angle, $\theta_{\rm app}$, defined from $z_{\rm cl}$ using Eq.~(\ref{eq:zclvstheta}), 
\begin{equation}
\theta_{\rm app}= \arcsin\left(1-\frac{1}{2}\left[ \frac{z_{\rm cl}}{l_\gamma}\right]^2\right)~.
\label{eq:thetaapp}
\end{equation}
As expected, this apparent contact angle decreases when the plate velocity is increased (squares, figure~\ref{fig: ThetaCa}a). However, $\theta_{\rm app}$ is far from zero at the entrainment transition, since $z_{\rm cl}$ remains well below the theoretical maximum of $\sqrt{2}l_\gamma$. Interestingly, our data for $\theta_{\rm app}$ can be directly compared with dynamic angle measurements for the same physico-chemical system, but for a different geometry, namely droplets sliding down an inclined plane (\cite{RDAL05}). Figure~\ref{fig: ThetaCa} shows that the two sets of data for a receding contact line are very similar, suggesting that the dynamic contact angle has some universal features. One should be careful, however, since \cite{RDAL05} measure the {\em actual} slope of the interface at a fixed distance from the contact line, while definition (\ref{eq:thetaapp}) represents an apparent slope when extrapolating static profiles. 

While the behavior of the dynamic contact angle appears to be robust with respect to the large scale geometry, the threshold {\rm Ca} for the entrainment transition is far from universal. In the experiments on sliding drops performed with the same substrate and liquids, the rear of the drop assumes a conical shape such that receding contact lines move at a constant normal velocity (\cite{PFL01,RDAL05}). The corresponding critical capillary number is  ${\rm Ca}_d = 5.7 \times 10^{-3}$, which is substantially lower than ${\rm Ca}^*= 9.1 \times 10^{-3}$. Yet another geometry gives a third different value: when the plate is pulled out at ${\rm Ca} > {\rm Ca}^*$ a liquid film is entrained except at the edges. As a result, a triangular (figure~\ref{fig: ThetaCa}b) or trapezoidal (figure~\ref{fig: ThetaCa}c) film is created. The receding speed of the lateral lines is constant and corresponds to ${\rm Ca}_>=7.5 \times 10^{-3}$. This shows that the threshold for contact line stability is not universal but depends on the details of the large scale geometry of the flow.
\begin{figure}
\includegraphics {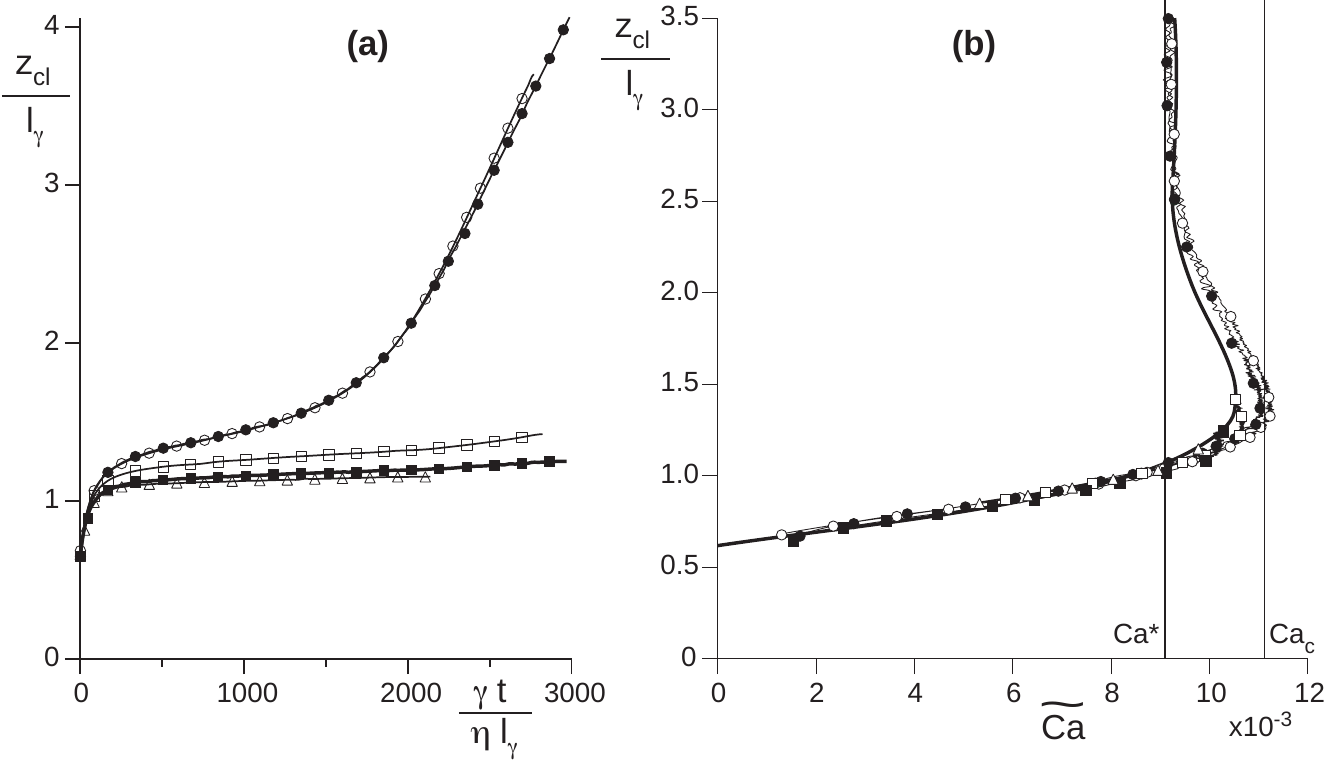}
\caption{{\em (a)} Meniscus height $z_{\rm cl}$ rescaled par the capillary length as a function of time, rescaled by the capillary time for ${\rm Ca}=9.7 \times 10^{-3}$ ($\triangle$), ${\rm Ca}=10.2 \times 10^{-3}$ ($\blacksquare$), ${\rm Ca}=10.7 \times 10^{-3}$ ($\square$), ${\rm Ca}=11.2 \times 10^{-3}$ ($\bullet$) and ${\rm Ca}=11.5 \times 10^{-3}$ (${\Large \circ}$). {\em (b)} Same data plotted as $z_{\rm cl}(t)$ as a function of the contact line relative capillary number $\widetilde{\rm Ca}=\eta (U_p-\dot{z}_{\rm cl}(t))/\gamma$. Solid line: Steady solutions of the multi-scale hydrodynamic model.}
\label{fig:bourrelet}
\end{figure}

\subsection{Experimental determination of ${\rm Ca}^*$}

The dynamical evolution from the steady meniscus to the ridge solution provides crucial information on the wetting transition. Figure~\ref{fig:bourrelet}a shows the time evolution of the meniscus height $z_{\rm cl}(t)$ after setting the plate velocity at a constant value at $t=0$. When ${\rm Ca} < {\rm Ca}^*$, $z_{\rm cl}$ relaxes exponentially to a nearly flat plateau. Note that we systematically observe a very slow upwards drift at a rescaled velocity $\eta \dot{z}_\infty/\gamma \sim 2 \cdot 10^{-5}$, which is three orders of magnitude smaller than typical capillary numbers. Above ${\rm Ca}^*$, the exponential relaxation is followed by a moderate steady rise and finally a much steeper rise corresponding to the development of the capillary ridge. Indeed, figure~\ref{fig:zcl}b shows that there is a well-defined point at which the contact line velocity exceeds the "noise" level, which allows to identify the entrainment transition. 

For ${\rm Ca} > {\rm Ca}^*$, liquid is entrained by the plate. As can be seen from the photograph of figure~\ref{fig: ThetaCa}c, the interface dynamics is not trivial: immediately behind the contact line we observe the formation of a capillary ridge. We have found experimentally that this structure travels exactly at a speed ${\rm Ca}^*$, suggesting that the threshold of entrainment is determined by properties of the ridge (\cite{SDFA06}). In fact, the ridge consists of two flat films that are connected through a capillary shock. On the one hand the boundary conditions at the contact line select a film thickness $h \propto l_{\gamma} {\rm Ca}^{*1/2}$, which is much thicker than the film connected to the bath, obeying the classical Landau-Levich scaling $h \propto l_{\gamma} {\rm Ca}^{2/3}$ (\cite{LL42}). This mismatch then gives rise to the shock. 

The picture that emerges is thus that, experimentally, entrainment occurs whenever the ridge can nucleate, even though stationary, linearly stable meniscus solutions in principle exist between ${\rm Ca}^*$ and ${\rm Ca}_c$. We believe that this avoided critical behavior is due to intrinsic noise in the experiment: contact angle hysteresis is a manifestation of microscopic inhomogeneity, an effect that is not treated in the model. The observation that ${\rm Ca}_c$ is sensitive to minor changes in the microscopic $\theta_{\rm cl}$, and the presence of contact line drift even below the transition support this interpretation.

\subsection{Quasi-steady transients: bifurcation diagram}

Let us now show how transient states during entrainment provide access to the full bifurcation structure of the wetting transition. The dynamical evolution towards a ridge can be recast in the plane $(z_{\rm cl}, \widetilde{{\rm Ca}})$, where $\widetilde{{\rm Ca}}$ is the capillary number based on the \emph{relative} velocity between plate and contact line, $U_p-\dot{z}_{\rm cl}$. Figure~\ref{fig:bourrelet}b represents parametric plots of $z_{\rm cl}(t)$ and $\widetilde{{\rm Ca}}(t)$, for different plate velocities. Surprisingly, all data points for various {\rm Ca} follow a single master curve. In addition, these points accurately follow the hydrodynamic prediction for the equilibrated values of $z_{\rm cl}$ versus ${\rm Ca}$ (solid line). Let us stress that this correspondence is far from trivial, since the theory considers stationary rather than dynamical interface profiles. Roughly speaking, one can identify (i) a stable branch ($d z_{\rm cl}/d{\rm Ca}>0$) on which all the steady menisci are located, (ii) an unstable branch ($d z_{\rm cl}/d{\rm Ca}<0$) where no steady menisci can exist, but where the we observe transients, and (iii) a vertical branch at ${\rm Ca} = {\rm Ca}^*$ corresponding to the velocity of the capillary ridge. The hydrodynamic theory predicts a slightly more complex structure with small oscillations around the vertical asymptote, which can not be resolved experimentally.

In addition to this correspondence, the data from the transient menisci can be compared to the values of $z_{\rm cl}$ for steady menisci obtained at ${\rm Ca} < {\rm Ca}^*$ (open circles, figure~\ref{fig:zcl}a). Indeed, the two data sets coincide, providing further evidence that transients states are similar in nature to the steady interface profiles.

These experimental findings strongly suggest that entrainment proceeds through a succession of steady states, which we refer to as a {\em quasi-steady} dynamics. Experimentally, the critical point (with a vertical tangent on the $z_{\rm cl}({\rm Ca})$ curve) is never reached through stationary menisci. However, \emph{during the transient} the meniscus shapes follow the complete bifurcation curve, and therefore provides an indirect measurement of ${\rm Ca}_c$. The critical capillary number is found here to be $11.1 \times 10^{-3}$, a slightly larger value than predicted by the hydrodynamic theory.
\begin{figure}
\includegraphics{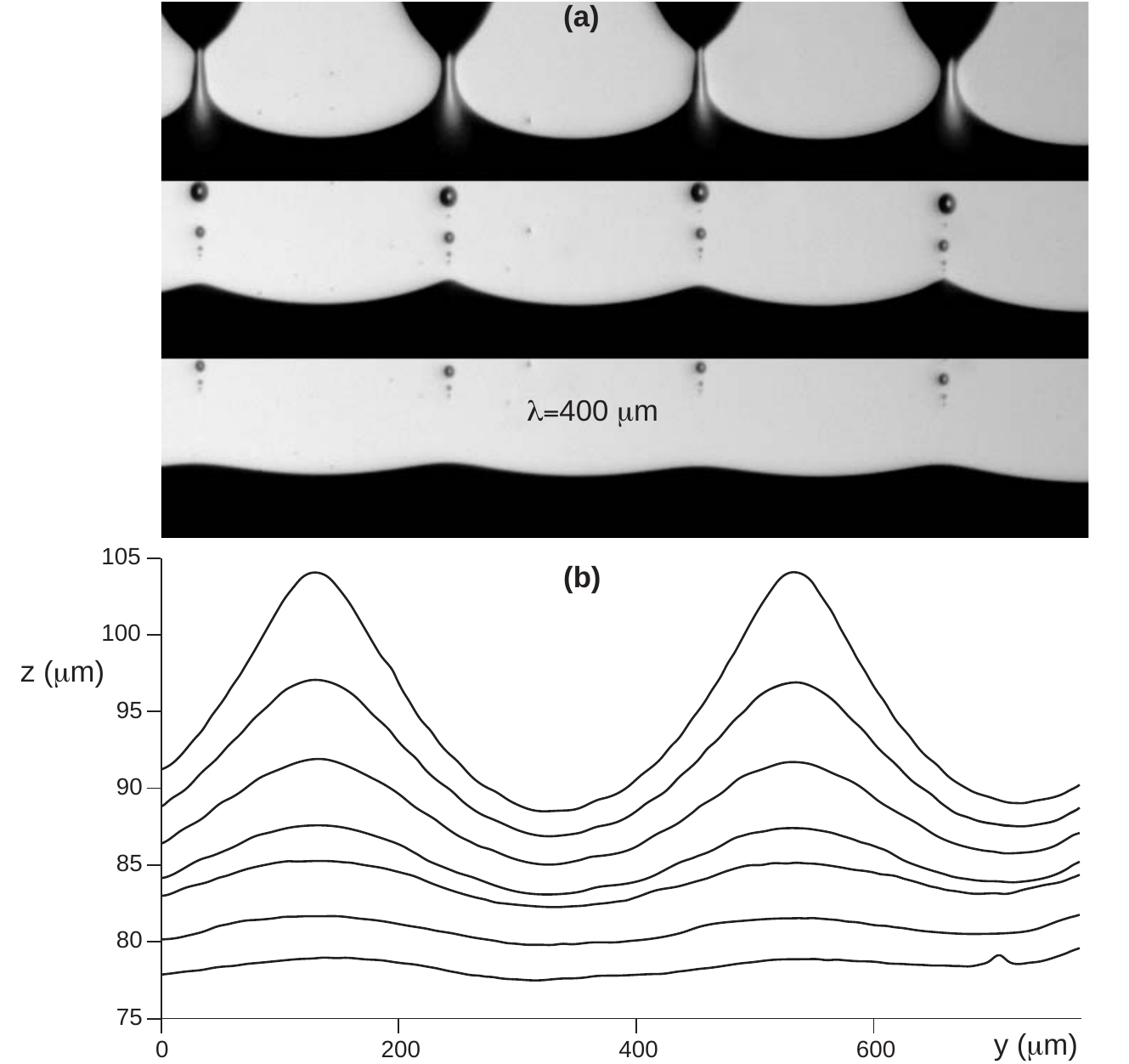}
\caption{{\em (a)} Pictures showing the evolution of the contact line initially perturbed at wavelength $\lambda=400~{\rm \mu m}$ by chemical defects on the plate. $\Delta t=0.4$~s. {\em (b)} Extracted profiles.}
\label{MultiDefautBrut}
\end{figure}

\section{Dispersion relation}\label{sec:dispersion}

Having discussed the dynamics of unperturbed menisci we can address perturbations of the contact line. As originally suggested by Golestanian \& Raphael, these should provide a sensitive experimental probe of small scale dynamics. In this section we consider two types of perturbations on the contact line: i) spatially periodic perturbations with rows of equally spaced defects (finite wavenumber $q$), ii) a global vertical shift of the meniscus ($q=0$). We first describe the experimental protocols, while the experimental findings results are discussed in the following section.

\subsection{Periodic defects}

To assess the dispersion relation, $\sigma$ versus $q$, as a function of the contact line speed, we performed systematic experiments with periodically spaced defects. A horizontal row of defects is created on the solid plate, as described in Sec.~\ref{sec:setup}. When this row of defects moves through the meniscus, it entrains drops of silicone oil out of the bath. As the defects move away from the meniscus, the threads connecting the drops to the bath pinch off leaving a few satellite droplets (figure~\ref{MultiDefautBrut}a). Immediately after the release from the defects, the contact line has a spatially periodic perturbation with sharp peaks, which decay quickly leaving a smoother almost sinusoidal perturbation. The spacing between defects is well below the capillary length, $\lambda = 400 \mu$m or $600 \mu$m, corresponding to $ql_{\gamma}\approx 23$ and $15$ respectively. As a consequence, the gravitational energy involved in the meniscus deformation is much smaller than the interfacial energy.
\begin{figure}
\includegraphics{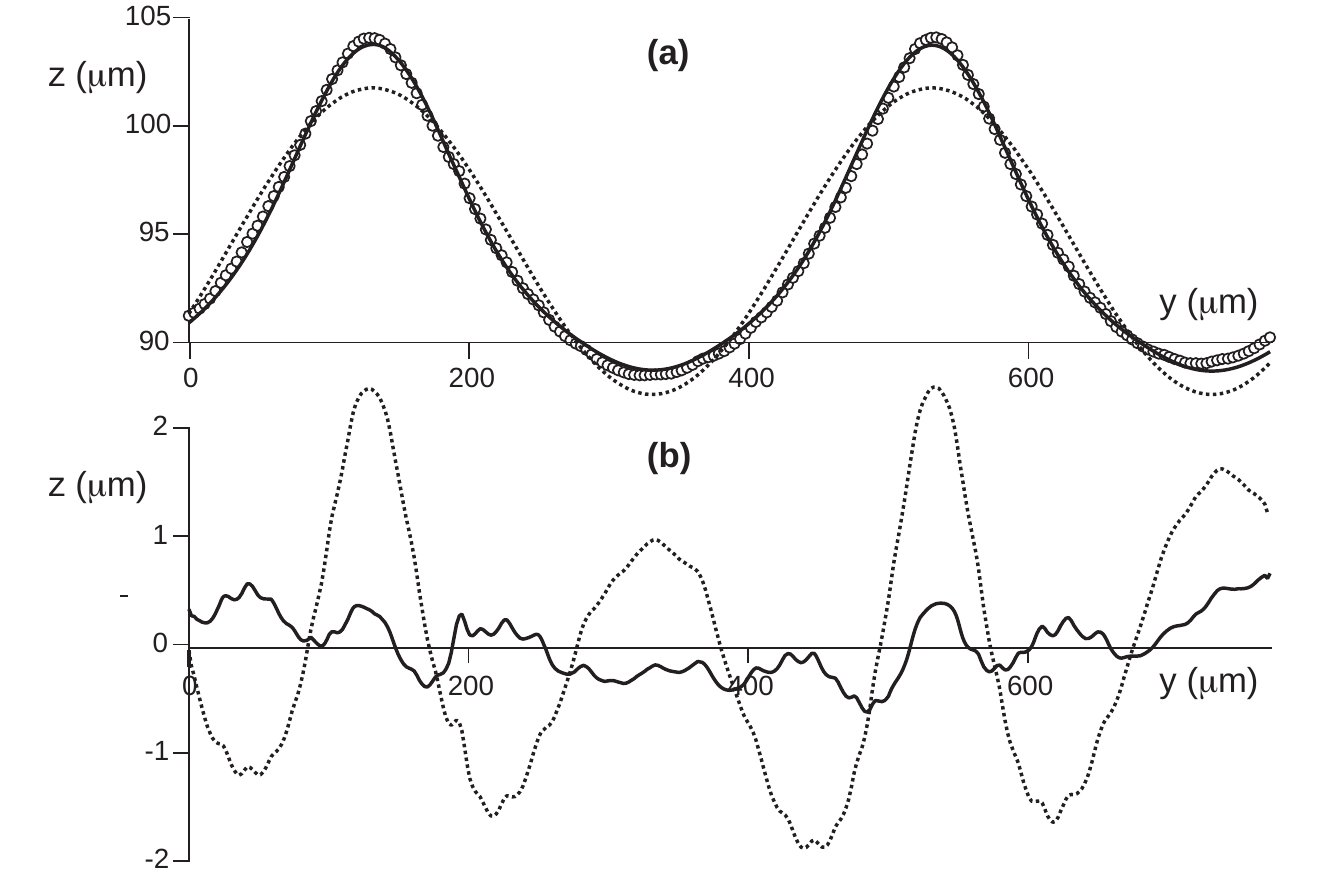}
\caption{{\em (a)} Fit of the contact line profile ($\circ$) by a single mode of wavelength $\lambda=400~{\rm \mu m}$ (dotted line) and by the sum of three modes, $\lambda=400~{\rm \mu m}$, $\lambda=200~{\rm \mu m}$ and $\lambda=133~{\rm \mu m}$ (solid line). {\em (b)} Corresponding residual ($z-z_{\rm fit}$) curves}
\label{MultiDefautFit}
\end{figure}

The precise location of the contact line is determined as described in Sec.~\ref{sec:setup} and the relaxation is analyzed over a horizontal range spanning two defects (see figure~\ref{MultiDefautBrut}b). Even if the defects are identical and evenly spaced, the liquid thread pinch-off generically do not occur simultaneously on all defects. For example, figure~\ref{MultiDefautBrut}a shows the pinch-off from four defects: on the top photograph, the rightmost liquid thread is clearly wider than the middle ones. It will then break slightly later. In the middle photograph, the corresponding peak is sharper and higher. Even after the decay of the highest spatial modes, there is still a small difference between peak amplitudes (figure~\ref{MultiDefautBrut}a, bottom photograph). For this reason it is impossible to fit the whole experimental curve with a single function and we choose to fit the curve by parts, considering only two defects at the same time (figure~\ref{MultiDefautBrut}b).

\begin{figure}
\includegraphics{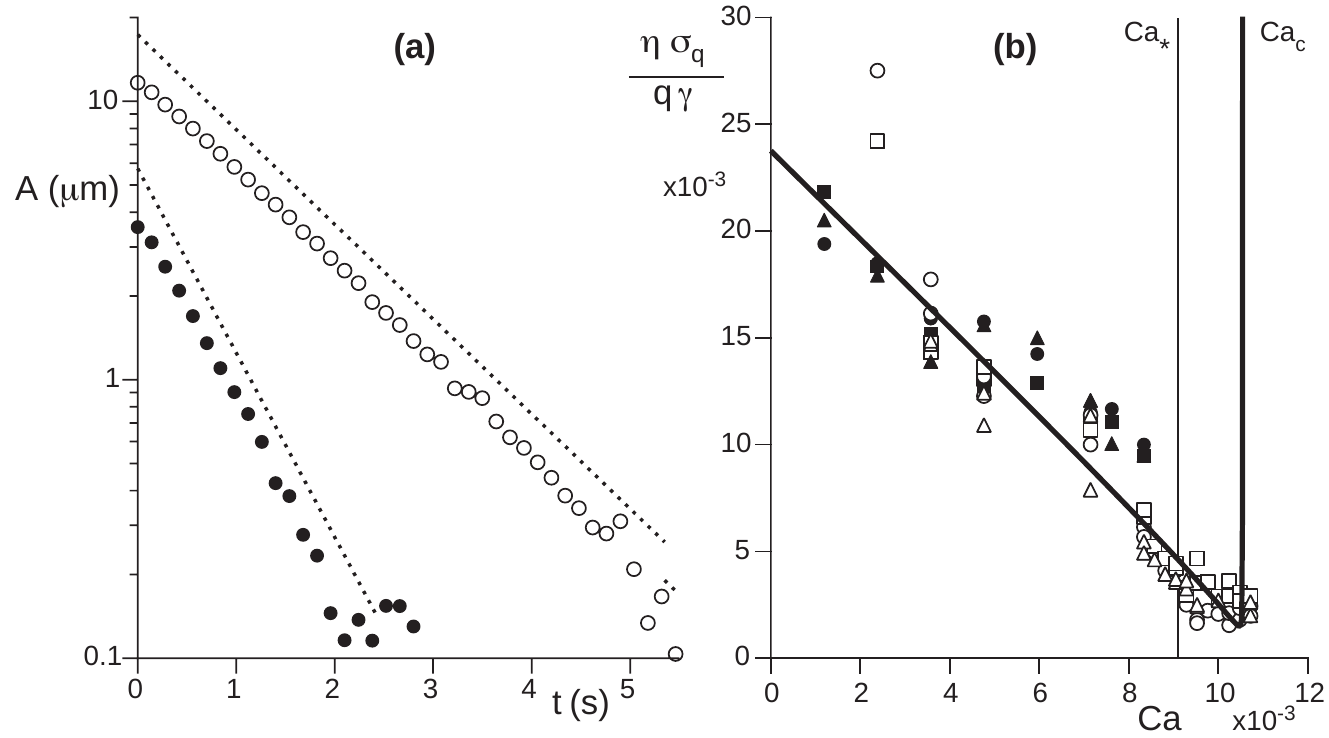}
\caption{{\em (a)} Amplitude of contact line deformation as a function of time for periodic perturbations. Open circles $\lambda = 400 \mu$m, filled circles $\lambda = 200 \mu$m. The dotted lines are exponentials. {\em (b)} Dimensionless relaxation rate as a function of capillary number  at different wavelengths (defects with 600 $\mu$m spacing: $(\circ) \lambda = 600 \mu$, $(\square)  \lambda = 300 \mu$, $ (\vartriangle)  \lambda = 200 \mu$ ; defects with 400 $\mu$m spacing: $(\bullet) \lambda = 400 \mu$, $(\blacksquare)  \lambda = 200 \mu$, $ (\blacktriangle)  \lambda = 133 \mu$ ). The solid line is the prediction of the multiscale hydrodynamic model with $\theta_{\rm cl}=(\theta_a+\theta_r)/2$.}
\label{fig:qmode}
\end{figure}

To analyze the relaxation, the experimental profiles are fitted by the sum of three modes: $z_{\rm fit} = a_0 + a_1 \cos (q_0 (y-\phi)) +  a_2 \cos (2 q_0 (y-\phi))  + a_3 \cos (3 q_0 (y-\phi)) $ where $q_0=2\pi/\lambda$ is the wavector corresponding to the spacing $\lambda$ between defects. It can be seen on figures~\ref{MultiDefautFit} that  a single cosine function does not fit the experimental curves correctly while the three mode fit gives an accurate description: for a total amplitude of 15 $\mu$m, the difference between the experimental points and the three mode fit is less than $0.5 \mu$m. We thus obtain the dynamics of three different wavevectors in a single experiment. This procedure allows a very precise determination of the amplitude (figure~\ref{fig:qmode}a), with a resolution exceeding the camera resolution. This is due to the averaging procedure which is implied by the fit over hundreds of data points.

For the three modes used in the fitting function, the amplitude decays exponentially as $e^{-\sigma t}$ (figure~\ref{fig:qmode}a), with a decay rate $\sigma$ proportional to the wavevector (mode 2 decays twice as fast as mode 1 and mode 3 three times faster than mode 1). As we will show below (figure~\ref{fig:dispersion}a), the data derived from the relaxation of multiple defects perturbation indeed display the linear relation between the relaxation rate $\sigma$ and the wavevector $q$, within experimental error, as anticipated in Eq.~(\ref{eq:scalingpgg}). 

\subsection{"Zero mode" relaxation}
\begin{figure} 
\includegraphics{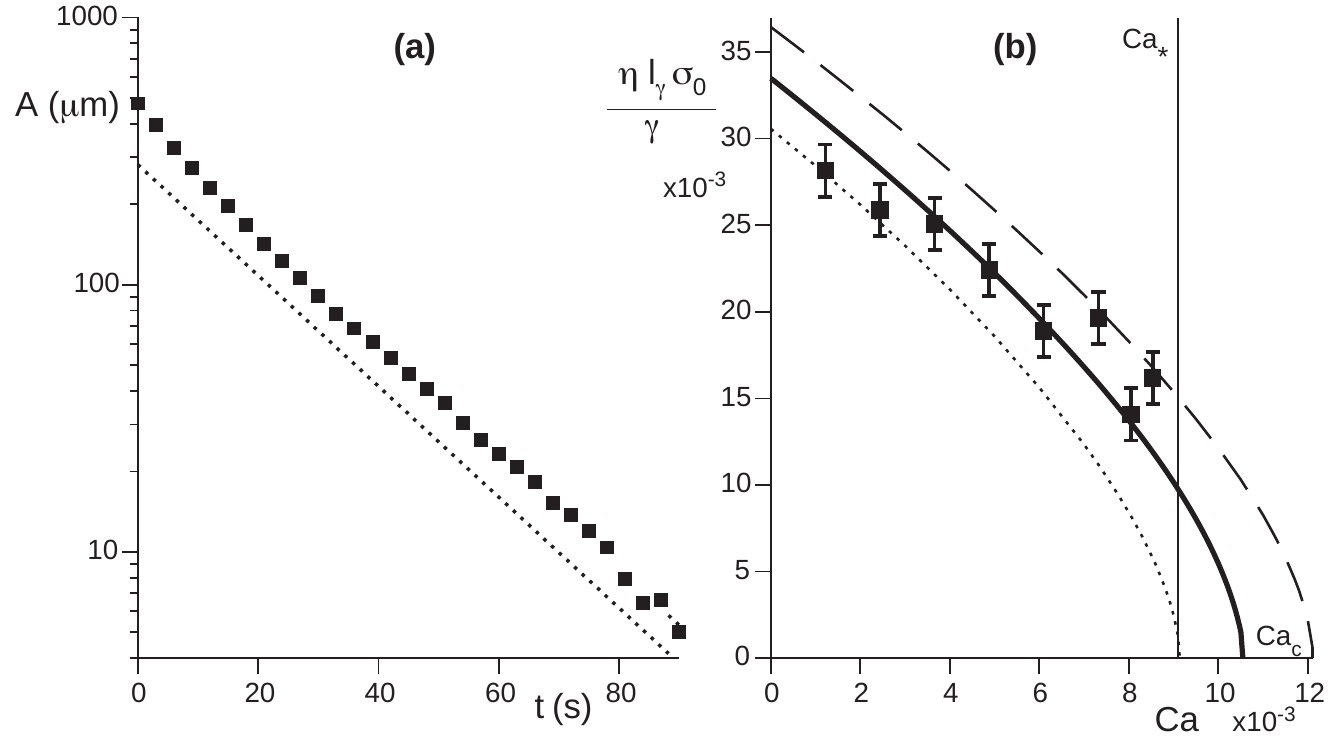} 
\caption{{\em (a)} Amplitude of contact line perturbation as a function of time for the "zero" mode $\lambda \rightarrow \infty$. {\em (b)} Dimensionless relaxation rate for the zero mode as a function of capillary number. The data have been obtained with the same plate. The error bars indicate the typical variation from one experiment to the other. The lines are the prediction of the multiscale hydrodynamic model for $\theta_{\rm cl} = \theta_r$ (dotted line), $(\theta_r+\theta_a)/2$ (solid line), $\theta_a$ (dashed line).} 
\label{fig:zeromode} 
\end{figure}

The experiments with regularly spaced defects provide data only in the long wavevector limit $ql_{\gamma} \gg 1$. But, we can get information on the small wavector limit $q \to 0$ simply by considering the relaxation of the meniscus height $z_{\rm cl}$ towards its steady value. Again, the amplitude of perturbation decays exponentially with time (figure~\ref{fig:zeromode}a). We fit the curves $z_{\rm cl}(t)$ for ${\rm Ca} < {\rm Ca}^*$ (as shown on figure~\ref{fig:bourrelet}a) by a function: $z_{\rm fit} = (z_{cl} + \dot{z}_{\infty}t)\left[1- e^{- \sigma t}\right]$, in which we account for the long term drift of the contact line through the term $\dot{z}_{\infty}t$. We thus obtain the relaxation rate $\sigma$ of the $q=0$ mode as a function of the capillary number. 

\section{Dimensionless relaxation rates and their evolution with {\rm Ca}}\label{sec:results}

We now analyze the experimentally measured relaxation rate, $\sigma$, as a function of $q$ and ${\rm Ca}$. In order to compare the obtained this experimental dispersion relation to theoretical predictions, we define dimensionless relaxation rates with different scalings in the limits $ql_{\gamma} \gg 1$ and $ql_{\gamma} \ll 1$. 

\subsection{Short wavelengths: $ql_\gamma \gg 1$}

Gravity plays no role in the large wavevector limit, so the only length scale in the problem is provided by the  wavelength of the perturbation. Hence, we expect the relaxation rate to scale with the imposed deformation $|q|$ and the characteristic capillary velocity $\gamma/\eta$ (\cite{DG86a}). We therefore introduce the dimensionless relaxation rate $\sigma_\infty({\rm {\rm Ca}})$:
\begin{equation}
\sigma = \frac {\gamma}{\eta } |q| \, \sigma_\infty({\rm Ca})~,
\label{eq:scalingwithq}
\end{equation}
where the subscript $\infty$ refers to the limit $q l_\gamma \rightarrow \infty$ (see also Part 1). 

The quasi-static theory for contact lines predicts $\sigma_\infty$ in terms of the apparent contact angle $\theta_{\rm app}$ and its dependence on ${\rm Ca}$ (\cite{GR03})
\begin{equation}
\sigma_\infty = -\theta \left(\frac{d \theta_{\rm app}}{d {\rm Ca}}\right)^{-1}~, 
\label{eq:sigmaqvsCa}
\end{equation}
which is the small angle limit of a more general expression. For all models of $\theta_{\rm app}({\rm Ca})$ (such as \cite{C86,V76,DG86b,B95}), $\sigma_\infty$ is found to decay almost linearly with ${\rm Ca}$, down to a zero value at the critical capillary number for entrainment. This implies a diverging relaxation time $\sigma^{-1}$, a direct consequence of the diverging slope $d\theta_{\rm app}/d{\rm Ca}$ at the critical point. The slope of the curve $\sigma_\infty({\rm Ca})$ varies from to -2 to -4, depending on the model used (\cite{GR01a}).

If we examine our experimental data (figure~\ref{fig:qmode}b), we can see that $\sigma_\infty$ indeed decreases almost linearly from ${\rm Ca}=0$ to ${\rm Ca}={\rm Ca}^*$, the location of the entrainment transition. But, this decreasing trend persists \emph{beyond} ${\rm Ca}^*$ when we consider the data points obtained during the transition. As we have shown in Sec.~\ref{sec:steady}, the transient meniscus adiabatically follows the bifurcation curve so we can effectively probe the contact line dynamics up to the critical point ${\rm Ca}_c$. The experiments clearly show that $\sigma_\infty$ does {\em not} go to zero between ${\rm Ca}^*$ and ${\rm Ca}_c$. This experimental result is in disagreement with the quasi-static theories. 

If, however, the viscous dissipation is accounted for in the full-scale hydrodynamic calculation, one indeed recovers a non-zero value of $\sigma_\infty$ at the critical point (Part 1, \cite{SADF07}). The prediction of hydrodynamic theory is represented by the solid line in figure~\ref{fig:qmode}b, where we took the microscopic contact angle as $\theta_{\rm cl}=(\theta_a+\theta_r)/2$. It correctly describes the variation of $\sigma_\infty$ over the whole range of capillary numbers, including the nonzero value at the critical point. Note that the solid line displays a sudden divergence near ${\rm Ca}_c$, which is due to a breakdown of the linear scaling $\sigma_q\propto |q|$ at criticality. This subtle effect is not observed within the experiments, for which the scaling with $|q|$ holds within experimental error. 

\subsection{Long wavelengths: $q l_\gamma \ll 1$}

In the small wavector limit, the energy of deformation is dominated by gravity and the relevant length scale is no longer provided by the wavelength, but the capillary length $l_\gamma$ (\cite{NV03}). We therefore define the dimensionless relaxation rate $\sigma_0({\rm Ca})$ as
\begin{equation}
\sigma = \frac {\gamma}{\eta l_{\gamma} } \, \sigma_0({\rm Ca})~.
\label{eq:scalingwithlcapp}
\end{equation}
The quasi-static theory predicts a dependence with ${\rm Ca}$ of the form 
\begin{equation}
\sigma_0 = l_\gamma \left(\frac{d z_{\rm cl}}{d {\rm Ca}}\right)^{-1}~,
\label{eq:sigma0vsCa}
\end{equation}
which was found in excellent agreement with the hydrodynamic calculation of Part 1. This relaxation is based on the idea that all transients with $q=0$ effectively obey a {\em quasi-steady} dynamics governed by a universal curve $z_{\rm cl}({\rm Ca})$, a concept that we discussed already in Sec.~\ref{sec:steady}. The critical point is again associated to a divergence of the slope $dz_{\rm cl}/d{\rm Ca}$, leading to a zero value of $\sigma_0$ at ${\rm Ca}_c$. In our experiments, we can only measure the relaxation towards a steady meniscus, i.e. when ${\rm Ca}$ remains smaller than ${\rm Ca}^*$. Within this limit, the model accounts reasonably well for the variation of $\sigma_0$.

\begin{figure}
\includegraphics{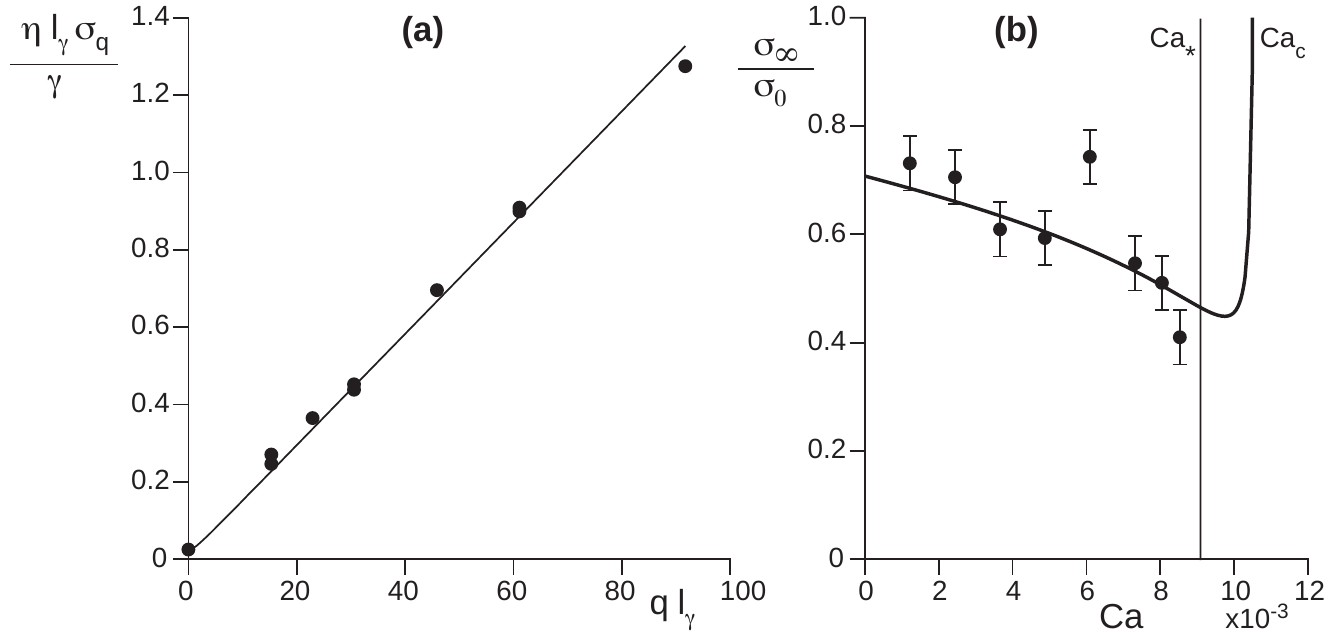} 
\caption{{\em (a)} Normalized relaxation rate as a function of the wavenumber rescaled by the capillary length. {\em (b)} Ratio of the relaxation rate of large wavenumber modes to zero mode rescaled by $ql_\gamma$, i.e. $\sigma_\infty/\sigma_0$. Each point corresponds to an average over several measurements. The error bars indicate the variance around the average. The hydrodynamic theory is presented by a solid line ($\theta_{\rm cl}=(\theta_a+\theta_r)/2$).} 
\label{fig:dispersion} 
\end{figure}

To close this section, let us compare the values of $\sigma_\infty$ and $\sigma_0$, by plotting their ratio in figure~\ref{fig:dispersion}b as a function of ${\rm Ca}$. We find a very good agreement with hydrodynamic theory (solid line). The ratio diverges at ${\rm Ca}_c$ since $\sigma_0 \rightarrow 0$ at ${\rm Ca}_c$, not accessible experimentally, while $\sigma_\infty$ remains finite.

\section{Localized perturbation and Green's function}\label{sec:lorentz}

Having confirmed the scaling $\sigma \propto |q|$ for short wavelengths, we can further investigate this "anomalous elasticity" of moving contact lines (\cite{JdG84,DG86a}, Golestanian \& Raphael 2001). An interesting consequence of this dispersion relation is that the corresponding Green's function is a Lorentzian: a localized perturbation of the contact line, created by a single defect passing accross the interface, should thus decay self-similarly according to a Lorentzian profile. The width (amplitude) is supposed to increase (decrease) linearly in time.

Suppose that, at time $0$, the contact line deformation is described by a Lorentzian of width $w_0$ and area $A$: 
\begin{equation}
z(y,0)=\frac{A}{\pi w_0}\frac{1}{1+ y^2/w_0^2}~,
\label{eq:lorentzian}
\end{equation}
with a peak amplitude $A/\pi w_0$. Its Fourier transform is
\begin{equation}
\hat{z}_q(0) = \frac{A}{\sqrt{2\pi}}   \exp( - |q| w_0)~.
\end{equation}
Using Eq.~(\ref{eq:scalingwithq}), we get the Fourier transform after relaxation during a time $t$ as 
\begin{equation}
\hat{z}_q(t) = e^{-\sigma t}\,\hat{z}_q(0) = 
\frac{A}{\sqrt{2\pi}} \exp \left( -|q| \left[ w_0+\frac{\gamma \sigma_\infty}{\eta} t \right]\right)~,
\label{eq:}
\end{equation}
which can be inverted to 
\begin{equation}\label{eq:boe}
z(y,t)=\frac{A}{\pi w(t)} \frac{1}{1+y^2/w(t)^2}~,
\end{equation}
where the width increasing linearly in time:
\begin{equation}\label{eq:width}
w(t) = w_0+\frac{\gamma \sigma_\infty}{\eta} t~.
\end{equation}

\begin{figure}
\includegraphics {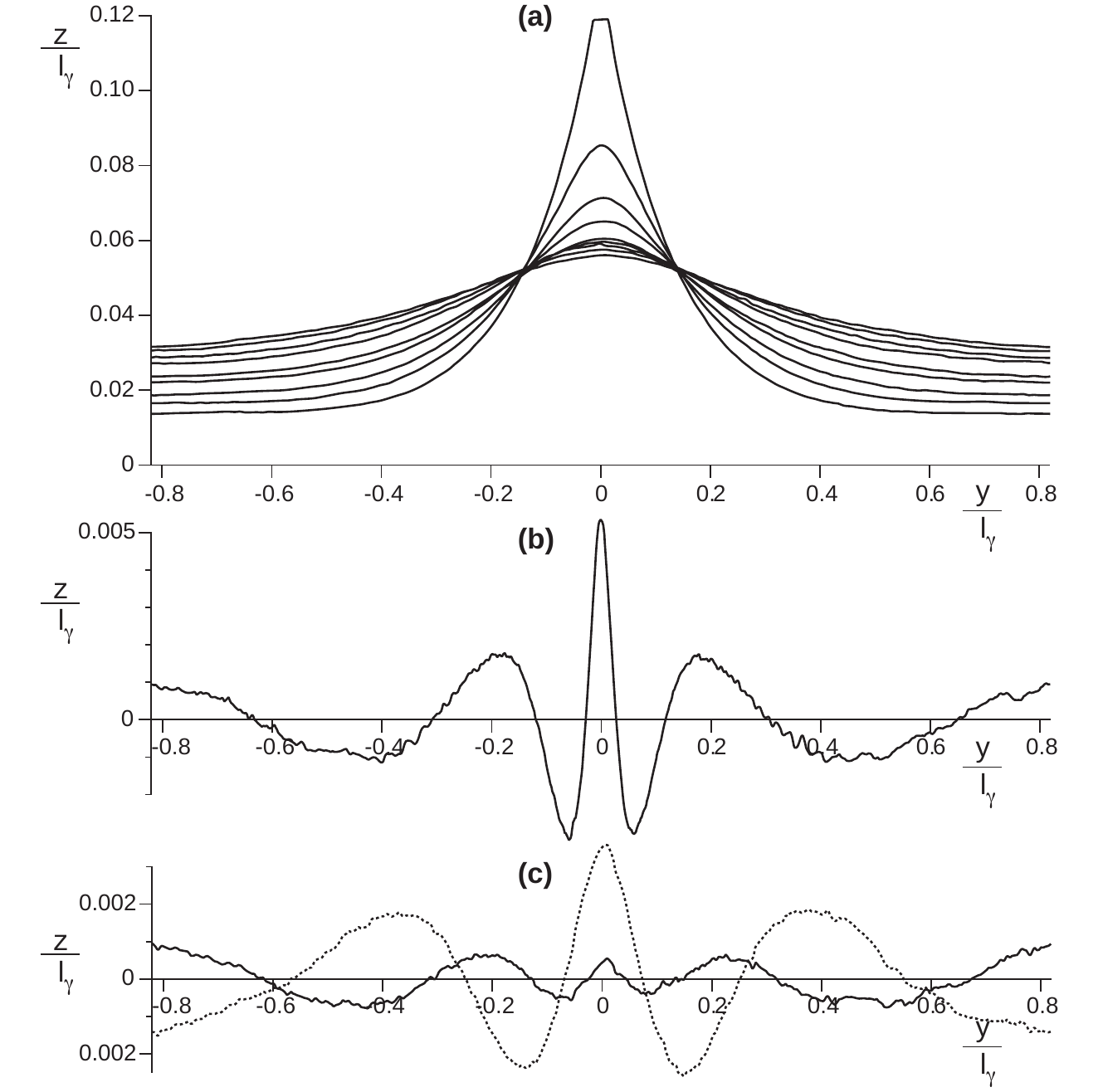}
\caption{{\em (a)} Time evolution of the contact line initially perturbed by a single chemical defect $\Delta t=2.5$~s, at ${\rm Ca} = 7 \times 10^{-3}$. {\em (b)} Residual of the fit of the contact line by a Lorentzian, just after depinning ($t=0.625$~s).  {\em (c)} Residual of the fit of the contact line by a Lorentzian (solid line) and by a Gaussian (dotted line) at time $t=5$~s.}
\label{MonoDefautBrut}
\end{figure}

Experimentally, we thus create a very localized perturbation that should quickly evolve into a Lorentzian shape. The time evolution of the perturbation created by a single defect is shown on figure~\ref{MonoDefautBrut}. In this experiment, the contact line speed is slightly below the critical speed. Immediately after depinning from the defect, the contact line is sharply peaked and cannot be fitted accurately by a Lorentzian (figure~\ref{MonoDefautBrut}b). After a few seconds, the modes corresponding to large wavenumbers are damped and the deformation is indeed very well approximated by a Lorentzian (for comparison we show a Gaussian fit in figure~\ref{MonoDefautBrut}c, dotted line). It is also worth noting that a logarithmic shape resulting from a localized force applied on the contact line (\cite{DG86a}) cannot describe properly the experimental profiles. 

\begin{figure}
\includegraphics {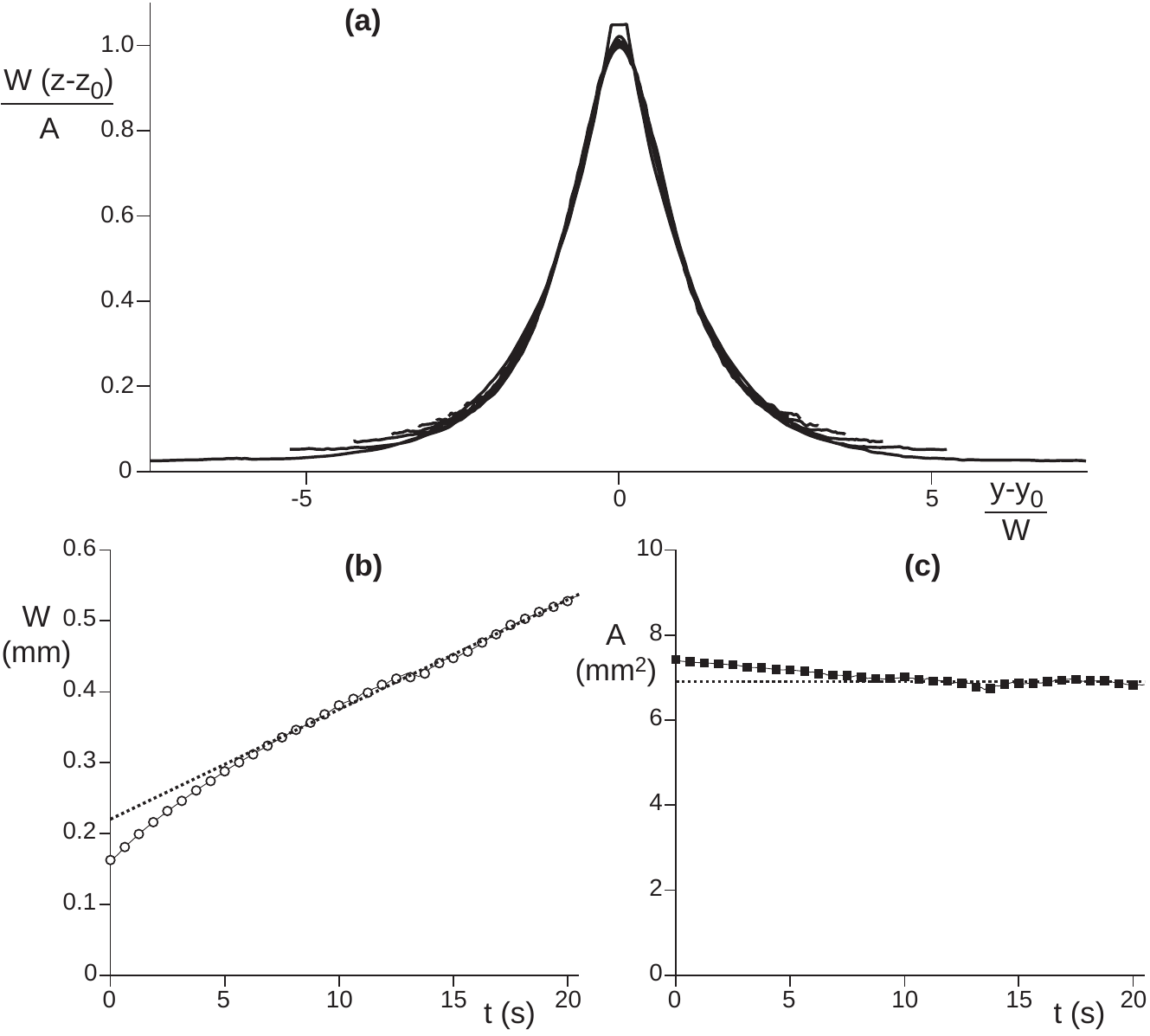}
\caption{{\em (a)} Rescaled contact line profiles (from figure~\ref{MonoDefautBrut}a) showing the self-similar behavior. {\em (b)}  Time evolution of the width derived from the fit. {\em (c)} Time evolution of the area $A$ under each curve.}
\label{MonoDefautCollapse}
\end{figure}

The convergence to a fixed Lorentzian shape is further evidenced by the rescaling of the experimental profiles $z(y,t)$, since Eq.~(\ref{eq:boe}) predicts $z(y,t)w(t)\pi/A(t) = f[y/w(t)]$. As expected, the shape of the rescaled curves nicely collapse onto a master curve, shown on figure~\ref{MonoDefautCollapse}a. Moreover, after the first few seconds during which the shape evolves into a Lorentzian, the computed width increases linearly with time (figure~\ref{MonoDefautCollapse}b). The spreading velocity of Eq.~(\ref{eq:width}), $\sigma_\infty \gamma/\eta$, was found to be $17 \mu$m/s in this example, corresponding to a dimensionless rate $\sigma_\infty \approx 8.4 \times 10^{-4}$. This value was obtained at $U = 140 \mu\mathrm{m/s}$ with 1 Pa.s oil, i.e. at $\mathrm{Ca} = 7 \times 10^{-3}$, very near the entrainment transition. The relaxation rate is indeed close to the lowest values observed with the periodic defects when ${\rm Ca}$ is between ${\rm Ca}^*$ and ${\rm Ca_c}$. Finally, the area under the fitting curve $A$ is found to be constant, again after the initial decay of the transient modes (figure~\ref{MonoDefautCollapse}c). 


\section{Conclusion}

We have measured the relaxation of a receding contact line, by considering perturbations in the limit of both small and large wavelengths with respect to the capillary length $l_{\gamma}$. This provides crucial information on the dynamics of contact lines and the nature of the dynamical wetting transition. As expected from the quasi-static theory by Golestanian \& Raphael, the moving contact line retains the peculiar elasticity already found for static lines, namely a relaxation rate proportional to the wavevector $q$, in the limit $q l_{\gamma} \gg 1$. However, their crucial prediction of diverging timescales at the entrainment transition is not confirmed experimentally. The initial interpretation for this was that the critical point is completely avoided through the nucleation of a capillary ridge (\cite{SDFA06}). However, the present experiments {\em do} explore the critical point through transients during liquid deposition: the interface profiles adiabatically proceed through stationary states, including the critical point. Even though, there is no evidence of a divergent relaxation time for perturbations of $ql_\gamma \gg 1$, which were found to decay on a very rapid time scale  even at criticality (figure~\ref{fig:qmode}b). 

These findings are consistent with the hydrodynamic calculation put forward in Part 1, in which we explicitly treat viscous effects at all length scales. There we showed that the critical point is described by a standard saddle-node bifurcation, for which $\sigma=0$ only for $q=0$, but not for finite wave perturbations. This demonstrates that a true hydrodynamic description is crucial to unravel the dynamics of contact lines. Another conclusion of Part 1 was that stationary menisci obey a rather surprising bifurcation diagram, that is characterized by two distinct capillary numbers, ${\rm Ca}^*$ and ${\rm Ca}_c$. The experimentally observed transients towards liquid deposition were indeed found to exhibit the same structure (figure~\ref{fig:bourrelet}b). 

There is, however, an important feature missing in the hydrodynamic description. Experimentally, the entrainment transition occurs at ${\rm Ca}^*$, while in theory solutions are linearly stable up to ${\rm Ca}_c$. \cite{S91} studied the entrainment transition for small siliconized glass rods pulled out of a bath of water-glycerin mixture. Within their experimental uncertainty, they found that entrainment occurs when the meniscus height is very close to its maximum value, with corresponding values of $\theta_{\rm app}$ ranging from 2 to 13$^\circ$ and this is in contradiction with our results. It should be noted that their substrates exhibit a large variation of static contact angle (from 70 to 86$^\circ$) and the magnitude of hysteresis is not reported. It is thus not clear if the discrepancy with our results is due to the strong interface curvature in the third dimension or to hysteresis effects.

A crucial step would be to incorporate substrate inhomogeneities into the theory. \cite{GR03} discussed the influence of fluctuations of surface energy (directly correlated to hysteresis) on the stability diagram for the wetting transition. They also predict, consistent with their quasi-static theory for smooth substrates, a roughening of the contact line at the coating transition since perturbations imposed by substrate heterogeneities should no longer relax. Our experimental and theoretical findings suggest a rather different scenario at the wetting transition, and underline the need for a hydrodynamic description incorporating hysteresis. 

Experimentally, it is extremely difficult to get rid of hysteresis on solid substrates. There have been attempts to use nanostructured surfaces: for example, \cite{SBVVEGdC00} used mixed alkanethiol monolayers to create composite surfaces with an hysteresis for alcane droplets varying from 2 to 7$^\circ$. They interpreted their results of droplet spreading (measuring an apparent contact angle as a function of time) in terms of the molecular kinetic theory of Blake (\cite{BH69}). They obtained a friction coefficient for the contact line which was correlated to the average composition of the thiol monolayer. As we have shown, dynamic characteristics near transitions are much more sensitive tests than quantities like apparent contact angles which are furthermore ambiguously defined. It will thus be interesting to perform experiments similar to those presented here, on substrates of viscous liquids to try to eliminate the hysteresis completely, or on nano-patterned solid substrates to try to vary the hysteresis continuously.

\begin{acknowledgments}
We wish to thank Elie Raphael who initially suggested this experiment. 
We also thank Jose Bico, Jens Eggers and Laurent Limat for fruitful discussions and Patrice Jenffer and David Renard for technical assistance. JHS acknowledges financial support by Marie Curie European Fellowships FP6 (MEIF-CT2003-502006, MEIF-CT2006-025104).
\end{acknowledgments}

\end{document}